\documentstyle[12pt,aaspp4,flushrt]{article}

\slugcomment{For ApJ, Part I}

\lefthead{Mould {\rm et~al. }}
\righthead{HST Key Project: The Hubble Constant}

\begin{document}
\vspace*{-0.9 truein}

\title{The $HST$ Key Project on the Extragalactic Distance Scale XXVIII. Combining 
the Constraints on the Hubble Constant
\footnote{Based on observations with the 
NASA/ESA \it Hubble Space Telescope\rm, obtained at the Space Telescope 
Science Institute, which is operated by AURA, Inc., under NASA Contract No. 
NAS 5-26555.}}

\author{Jeremy R. Mould\altaffilmark{2}, 
John P. Huchra\altaffilmark{3},
Wendy L. Freedman\altaffilmark{4},
Robert C. Kennicutt, Jr.\altaffilmark{5},
Laura Ferrarese\altaffilmark{6}, 
Holland C. Ford\altaffilmark{7}, 
Brad K. Gibson\altaffilmark{8},
John A. Graham\altaffilmark{9},
Shaun M.G. Hughes\altaffilmark{10},
Garth D. Illingworth\altaffilmark{11}, 
Daniel D. Kelson\altaffilmark{9}, 
Lucas M. Macri\altaffilmark{3}, 
Barry F. Madore\altaffilmark{12},
%Abhijit Saha\altaffilmark{13}, 
Shoko Sakai\altaffilmark{13},
Kim M. Sebo\altaffilmark{2},
Nancy A. Silbermann\altaffilmark{12} and
Peter B. Stetson\altaffilmark{14}, 
}

\altaffiltext{2}{Research School of Astronomy \& Astrophysics, Institute of Advanced Studies, Australian National University, Mount Stromlo Observatory, Weston, ACT, Australia  2611}
\altaffiltext{3}{Harvard-Smithsonian Center for Astrophysics, 60 Garden St., Cambridge, MA  02138}
\altaffiltext{4}{The Observatories, Carnegie Institution of Washington, Pasadena, CA, USA  }
\altaffiltext{5}{Steward Observatory, Univ. of Arizona, Tucson, AZ,   85721}
%\altaffiltext{7}{European Southern Observatory, D-85748 Garching b. M\"unchen, Germany}
\altaffiltext{6}{Hubble Fellow, California Institute of Technology, Pasadena, CA,   91125} 
\altaffiltext{7}{Dept. of Physics \& Astronomy, Bloomberg 501, Johns Hopkins Univ., 3400 N. Charles St., Baltimore, MD,   21218}
\altaffiltext{8}{CASA, University of Colorado, Boulder, CO 80309-0440}
\altaffiltext{9}{Dept. of Terrestrial Magnetism, Carnegie Institution of Washington, 5241 Broad Branch Rd. N.W., Washington, D.C.,  20015}
%\altaffiltext{11}{Avanti Corporation, 46871 Bayside Parkway, Fremont, CA, 94538}
%\altaffiltext{12}{Univ. of Wisconsin, Madison, WI, USA, 53706}
\altaffiltext{10}{Royal Greenwich Observatory, Madingley Road., Cambridge, UK  CB3~0EZ \newline
Current address: Institute of Astronomy, Madingley Road., Cambridge, UK  CB3~0HA}
\altaffiltext{11}{Lick Observatory, Univ. of California, Santa Cruz, CA, 95064}
\altaffiltext{12}{Infrared Processing and Analysis Center, Jet Propulsion Laboratory, California Institute of Technology, Pasadena, CA,  91125}
\altaffiltext{13}{National Optical Astronomy Observatories, P.O. Box 26732, Tucson, AZ  85726}
\altaffiltext{14}{Dominion Astrophysical Observatory, Herzberg Institute of
Astrophysics, National Research Council, 5071 West Saanich Rd., Victoria,
BC, Canada  V8X~4M6} 

%BEGIN USER-DEFINED SHORTFORMS
\def\spose#1{\hbox to 0pt{#1\hss}}
\def\simlt{\mathrel{\spose{\lower 3pt\hbox{$\mathchar"218$}}
     \raise 2.0pt\hbox{$\mathchar"13C$}}}
\def\simgt{\mathrel{\spose{\lower 3pt\hbox{$\mathchar"218$}}
     \raise 2.0pt\hbox{$\mathchar"13E$}}}
\def\eg{{\rm e.g. }}
\def\ie{{\rm i.e. }}
\def\etal{{\rm et~al. }}
\def\kms{{km~s$^{-1}~$}}
\def\hunit{{km~s$^{-1}~{\rm Mpc}^{-1}~$}}
%END USER-DEFINED SHORTFORMS

\begin{abstract}

Since the launch of Hubble Space Telescope nine years ago Cepheid distances
to 25 galaxies have been determined for the purpose of calibrating
secondary distance indicators. Eighteen of these have been measured by
the HST Key Project team, six by the Supernova Calibration Project,
and one independently by Tanvir.
Collectively this work sets out an array of survey markers over the region
within 25 Mpc of the Milky Way. A variety of secondary distance indicators
can now be calibrated, and the accompanying four papers employ the full
set of 25 galaxies to consider
the Tully-Fisher relation, the fundamental plane of elliptical galaxies,
Type Ia supernovae, and surface brightness fluctuations.

When calibrated with Cepheid distances, each of these methods yields
a measurement of the Hubble constant and a corresponding measurement 
uncertainty. We combine these measurements in this paper, together
with a model of the velocity field, to yield the
best available estimate of the value of H$_0$ within the range of these
secondary distance indicators and its uncertainty. The uncertainty
in the result is modelled in an extensive simulation we have called
``the virtual key project." The velocity field model includes the influence
of the Virgo cluster, the Great Attractor, and the Shapley supercluster,
but does not play a significant part in determining the result.

The result is H$_0$ = 71 $\pm$ 6 \hunit. The largest contributor to the
uncertainty of this 67\% confidence level result is the distance of the
Large Magellanic Cloud, which has been assumed to be 50 $\pm$ 3 kpc.
This takes up the first 6.5\% of our 9\% error budget.
Other contributors are the photometric calibration of the WFPC2 instrument,
which takes up 4.5\%, deviations from uniform Hubble flow in the volume 
sampled ($\simlt$2\%), the composition
sensitivity of the Cepheid period-luminosity relation (4\%), and departures
from a universal reddening law ($\sim$1\%). 
These are the major components, which, when combined in quadrature,
make up the 9\% total uncertainty. If  the  LMC distance modulus were
systematically  smaller by 1$\sigma$ than that adopted here,
the derived value of the  Hubble constant would increase by 4 \hunit.
Most of the significant systematic errors are capable of 
amelioration in future work.  These include the uncertainty in
the photometric calibration of WFPC2, the LMC distance, and the reddening
correction.
A NICMOS study is in its preliminary reduction phase, addressing the last of these.

Various empirical analyses have suggested that
Cepheid distance moduli are affected by metallicity differences.
If we adopted the composition sensitivity obtained in
the Key Project's study of M101, and employed the oxygen abundances
measured spectroscopically in each of the Cepheid fields we have studied,
the value of the Hubble Constant would be reduced by 4 $\pm$ 2 \% to
68 $\pm$ 6 \hunit. 

\end{abstract}


\section{Introduction}
\label{introduction}

The goal of the \it Hubble Space Telescope (HST) Key Project on
the Extragalactic Distance Scale \rm was announced in 1984 
by the newly formed Space Telescope Science Institute to be
the determination of the Hubble Constant to an accuracy $\simlt 10$\%. 
The recommended approach was classical: to use Cepheid distances to
calibrate secondary distance indicators. A plan was developed (Aaronson \&
Mould 1986), and our proposal was selected, but the project did not get into 
top gear until HST's spherical aberration had been corrected.
This observing program, which is now complete, has been described in detail by Kennicutt, Freedman \& Mould (1995).  

The accompanying four papers (Sakai \etal 2000; Gibson \etal 2000; Kelson
\etal 2000; Ferrarese \etal 2000a) show how Cepheid
distances to 18 spirals, within $\sim 25$ Mpc, are used to calibrate the
Tully-Fisher relation for spiral galaxies (TF), 
the fundamental plane for elliptical galaxies (FP),
surface brightness fluctuations (SBF), 
and (with the 6 additional galaxies of the SN calibration project)
Type Ia supernovae. Each of these distance indicators is able to penetrate
to sufficient distance (10$^4$ \kms) that perturbations in the Hubble flow are
small compared with the expansion of the Universe.
None of them is free from implicit assumptions about the stellar population
of the galaxies whose distances are being measured.
That is the case, whether we are assuming constancy of mass-to-light ratio
between the galaxies whose Cepheid distances we have measured 
(``the calibrators") and galaxies in, say, the Coma cluster, or whether
we are assuming that supernova progenitors are essentially similar 
in the calibrators and the Cal\'an/Tololo survey galaxies.
It therefore seems more prudent to combine constraints from four separate
secondary distance indicators with different systematics, than to 
investigate one alone.
 
The purpose of the present paper is to show how these constraints on H$_0$
can be combined to yield the local Hubble Constant to an accuracy $\simlt 10$\%
to 1-$\sigma$ confidence level. In a subsequent paper (Freedman \etal 2000) 
we examine the extrapolation to a sufficiently larger volume as to step
from a local value of H$_0$ to the global expansion rate.

\section{The Key Project Distance Database and Velocity Field Model}

The primary product of the Key Project has been the discovery of Cepheids
in a set of galaxies within 25 Mpc and the measurement of their characteristics.
The galaxy distances inferred by means of period luminosity relations are
collected by Ferrarese \etal (2000b). Secondary distance indicators extend
the range of measurement into the redshift range (2000, 10000) \kms, 
and it is then necessary to relate the recession velocities of these objects
to the smooth Hubble flow.

One of the major remaining uncertainties in the determination
of the Hubble Constant is the correction of the observed velocities
of our tracers for large scale motions.  Twenty years ago,
the ``cosmic'' velocities of objects were simply taken to be
their velocities corrected for galactic rotation and sometimes
additionally corrected to the centroid of the Local Group.
Slightly more than 20 years ago, the apparent motion of the Milky
Way and, by inference, the Local Group with respect to the Cosmic Microwave
Background (CMB) was discovered,  and it has now been exquisitely measured
with COBE (Kogut \etal 1993). Slightly less than 20 years ago, the 
infall of the Local Group
into the core of the Local Supercluster that had been predicted by 
de Vaucouleurs (1958; 1972), Peebles (1976), Silk (1974) and others was detected
(Tonry \& Davis 1980; Davis \etal 1980; Aaronson \etal 1982a; Aaronson \etal 1980).  Soon after, larger scale flows were seen (Burstein \etal 1986; 
Lynden-Bell \etal 1988).  

It is now clear (c.f. Strauss 
\& Willick 1995) that there are motions on scales of tens of Mpc with 
amplitudes up to of order the Milky Way's motion with respect to the CMB.  However,
the exact nature of these motions with respect to the CMB is still unclear
(Lauer \& Postman 1994; Riess, Press \& Kirshner 1995), as are
the precise causes of our motion.

Recently, several groups have chosen to treat this problem by
correcting apparent velocities to the CMB frame. This is generally
done by just applying a correction that is our CMB velocity
($\sim$630 \kms) times the cosine of the angle between the
direction of motion with respect to the CMB.  At large velocities, cz $\geq$
10,000 \kms, unless there are peculiar velocities on much larger 
scales or with much larger amplitudes than hitherto seen, this 
correction is both small and probably proper.  Not correcting
for it, in fact, can introduce a bias in the determination of
H$_0$ which could be as large as  V$_{CMB}$/V$_{object}$, or 6\%.
In fact, for any sample of objects  not uniformly distributed 
w.r.t. $\cos \theta_{CMB}$, such bias  could be a non-negligible
contribution to the error in H$_0$.

Worse than that, at smaller redshifts, the flow field is
much more complicated (c.f. Dekel \etal 1999); near the centers
of rich clusters the infall amplitudes and/or the velocity
dispersion can be very large ($\simgt$1,000 \kms) and for nearby objects
peculiar motions can be a substantial part of the observed
velocity.  It is also clear that it is a mistake to correct
the velocities of very nearby objects by a simple $\cos \theta_{CMB}$
term, because nearby objects (\eg M31 or the M81 group) are closer to
being at rest with respect to the Local Group frame than to the CMB frame.

To treat this problem for our various determinations of H$_0$
via several different samples of groups, clusters and individual
galaxies we have developed a simple linear multiattractor model based on 
the Han and Mould (1990) and Han (1992) models, and  similar 
to the multiattractor model advocated by Marinoni \etal (1998a).
The  model is linear and assumes (1) a flow towards each attractor
(\eg Virgo, the Great Attractor) that is independent of each object
so the corrections for each are additive, (2) flows described by a 
fiducial infall velocity at the position of the Local Group towards each 
attractor (c.f. Peebles 1976; Schechter 1980), and (3) an essentially 
cylindrical (section of a cone) masked volume around each 
attractor where objects are forced to the attractor's velocity.
This last procedure collapses the cluster cores and avoids
our having to deal with regions where the flow field is certainly non-linear
and usually multi-valued for any observed velocity. We add one additional
simplifying assumption, (4) to first order peculiar velocities are
small enough so that an object's apparent velocity in the Local Group
frame is the estimate  of its distance.  Again, this assumption is generally 
justified for objects far from our attractors. With these assumptions, it is trivial to include
additional attractors (\eg the Shapley Supercluster, Scaramella \etal 1989)
if desired.

The simple linear infall model has been described by a number of authors,
most notably Schechter (1980).  In this model, the estimated radial component
(with respect to the Local group) of peculiar velocity induced by an 
attractor is, by the law of cosines,

$$ V_{infall} \ \approx \ {V_{fid} \cos \theta} \ + \ V_{fid} \biggl({ {V_o - 
V_a \cos \theta}
\over {r_{oa}} }\biggr) \biggl({{r_{oa}} \over {V_{a}}} \biggr)^{1-\gamma}
\qquad (1)  ,$$

\noindent where $V_{fid}$ is the amplitude of the infall pattern to that attractor
{\it at the Local group},   $V_o$ is the observed velocity of the object
(in the LG frame),
$V_a$ is the observed distance of the attractor expressed as a velocity,
$\gamma$ is the slope of the attractor's density profile, $\rho(r) \ \propto
\ r^{-\gamma}$,
$\theta$ is the projected angle between the object and the attractor,
and $r_{oa}$ is the estimated distance of the object from the attractor expressed
as a velocity,
$$r_{oa}  \ = \ \sqrt{V_o^2 + V_a^2 -2V_o V_a \cos \theta} \qquad (2).$$
\noindent The first term in equation (1) is the projection of the LG 
infall velocity into the attractor ($V_{fid}$), and the second term is the 
projection of the object's infall into the attractor.

Note that we have modified the normal form of this relation, which
uses the true relative distances of the objects in question, to instead
express distances as velocities.  To produce our simple flow
field corrections, rather than solve for the actual relative distances
of the objects in question, we have assumed that, to first order, the 
apparent radial   velocity of an object (in the Local Group frame) 
represents its distance.  We fix and use the cosmic velocity of the
attractors, after solving for their own motions with respect to the other attractors.
A more complete treatment would iteratively solve for the true velocity
of each source, and an even more complete treatment would be to use
the actual observed density field (Marioni \etal 1998b), but since our
goal is to provide just a first order flow field correction to investigate
and eliminate significant flow field biases in our H$_0$ determinations,
we stop here.  The details are given in Appendix A.

This seems reasonable given the uncertainty in the absolute 
distance, and location for the main attractors, our significant lack of
knowledge of the flow field at distances much beyond 4500 km/s and
the other simplifying assumptions that are generally made such as assuming
spherical attractors. We will test the above assumption in future work
(Huchra \etal 2000),
and implement an iterative solution to the flow field corrections 
if it is warranted.  For the present, we note that if we abandoned our
flow field model and assumed instead that only the observer was in motion
relative to the smooth Hubble flow, the maximum change in our result
for H$_0$ would be a 4\% increase in the SBF result. There would be a 2\%
decrease in the result from supernovae and smaller effects from TF and FP.

\section{The Virtual Key Project}
\label{simulations}

The cosmic distance ladder  is a notable example of the
concatenation of measurement uncertainties in a multi-step experiment
(Rowan Robinson 1986).
Careful distinction between random and systematic error is required
(\eg Madore \etal 1998), and bias is a concern (Sandage 1996). For the
purposes of this paper, we have developed a simulation code which recreates
the Key Project in the computer, allowing the uncertainties and parameter
dependences to be followed  extensively and investigated rigorously.

One fundamental assumption of the Key Project is that the distance of the
Large Magellanic Cloud is 50 kpc (m-M = 18.50 $\pm$ 0.13 mag). Indeed,
our result might best be expressed in units of km/sec/LMC-distance.
Nevertheless, a 6.5\% uncertainty in the distance of the LMC is incorporated
in the project error budget. In $\S$5 we also explore
use of a literature survey (Figure 1) as a probability distribution function
for the LMC distance. Westerlund's (1996) survey has been updated, as discussed
in more detail by Freedman \etal (2000).

The LMC Cepheid period
luminosity (PL) relation in the simulation also has a 0.02 mag zeropoint
uncertainty (Madore \& Freedman 1991; Tanvir 1997). This estimate will be tested
when photometry of a larger sample of LMC Cepheids is
complete (Sebo \etal 2000). A second systematic error in the
Key Project arises from the residual uncertainty of WFPC2's
correction for Charge Transfer Efficiency (see Appendix B) and
calibration on to the (V,I) system. 
This is amplified to a 0.09 mag uncertainty in distance modulus
by the approach we have adopted to reddening correction, because each galaxy's
absolute distance modulus is a linear combination of the apparent moduli:
$\mu_0$ = 2.45 $\mu_I$ -- 1.45 $\mu_V$.

The metallicity dependence of the PL relation has proved  difficult
to constrain (Kennicutt \etal 1998). For most of the galaxies for which
we have measured Cepheid distances, however, measurements of oxygen 
abundances in HII regions in or near the Cepheid fields are also available.
Sakai \etal (2000), Gibson \etal (2000), Kelson
\etal (2000) and Ferrarese \etal (2000a) present results, both neglecting
PLZ, and also correcting the galaxy distances published in papers I--XXI
by the coefficient $\gamma_{VI}$ = d(m-M)/d[O/H]  = --0.24 $\pm$ 0.16 mag/dex
(Kennicutt \etal 1998),
and this is followed in the simulation. In each simulation a value of $\gamma$ is drawn from
a normal distribution for this purpose. Normal distributions are employed
throughout these simulations, except where otherwise noted.

For each of the galaxies in the virtual key project a Cepheid distance
is generated assuming a 0.05 mag intercept
uncertainty in its PLV relation and a similar intercept uncertainty in PLI.
These are typical values; some of the real galaxies have more Cepheids
and better determined distances (\eg NGC~925), and others fewer
Cepheids (\eg NGC~4414), and hence larger uncertainties.
Implicitly, we have assumed that the reddening law is universal, adopting
an uncertainty in its slope: R$_V$ = 3.3 $\pm$ 0.3.
As a follow-on to the Key Project, this will be tested with NICMOS 
observations of HST Cepheids for some galaxies.

A primary, and long awaited (Aaronson \& Mould 1986) outcome
of the Key Project is a calibration of the TF relation.
Sakai \etal (2000) find an $rms$ scatter about this relation,
and this is included in the 18 galaxy calibration simulation. The calibration
is then applied to a sample of 5 clusters with $cz$ $>$ 5000 \kms.
(And this is all then realized half a million times.)
Sakai \etal analyze a larger sample than this, but their final result
is based on the most distant members of the cluster dataset. 
Comparison of the simulated and input Hubble
relations yields an H$_0$ error from the TF calibration component of the
Key Project.

In the simulation
velocities are drawn from a normal distribution with a $\sigma$ = 300
\kms (Giovanelli \etal 1998). The real flow field is more complex,
and the model adopted by Sakai \etal (2000), Gibson \etal (2000), Kelson
\etal (2000) and Ferrarese \etal (2000a) is specified in Appendix A.
We have incorporated the Tully-Fisher error budget given by Sakai \etal
in their Table 6.
Noting the discrepancy they report between cluster distances
based on I band photometry and those based on H band photometry,
we adopt an uncertainty of 0.18 mag to allow for systematics in the galaxy
photometry.

Second, the Key Project provides a very direct calibration of the SBF relation.
Following Ferrarese \etal (2000a), the simulation takes six galaxies
and derives a zeropoint for the relation between SBF magnitude and color.
This relation, which is assumed to have an $rms$ scatter of 0.11 mag
(Tonry \etal 1997), is then applied to the four galaxies 
in the redshift range 3000 -- 5000 \kms with HST Planetary
Camera SBF measurements. We have omitted Coma and NGC~4373, just as Ferrarese
did. Comparison of the simulated and input Hubble
relations yields an H$_0$ error from the SBF calibration component of the
Key Project. The approach here follows the error budget in Table 5 of
Ferrarese \etal (2000a).

The third, and in some respects strongest, component of the distance scale
calibrated by HST's Cepheid database is the relation between
the maximum luminosity of the SNIa light curve and the supernova decline
rate (Hamuy \etal 1996; Phillips \etal 1999). 
Uncertainties in the observed magnitudes and reddening of the
supernovae and their Cepheid distances affect the calibration.
In the simulation we have adopted the uncertainties quoted by Gibson \etal
(2000) for six calibrators and incorporated the error budget given in their 
Table 7. The calibration was then applied to the 27 supernovae
of Hamuy \etal between 6,000 and 30,000 \kms.

Finally, to simulate the calibration of the fundamental plane we have 
have employed the error budget in Table 3 of Kelson \etal and assumed
that the Leo ellipticals lie within 1 Mpc of
the respective mean distances of their Cepheid-bearing associates.
In the case of Virgo and Fornax we have assumed elongation of the cluster
along the line of site, described by Gonzales \& Faber (1997) as an
exponential fall-off with a 2.5--4 Mpc scale length. The  calibration
is then applied to 8 clusters, ranging from Hydra to Abell 3381 in distance.
We have assumed that the clusters have the same 300 \kms $rms$
noise that is seen in the TF sample. 

\section{The Error Distributions}

Based on 5 $\times$ 10$^5$ realizations,
Figure 2a shows that the simulated TF error distribution, which gives the
probability that H$_0$ determined by Sakai \etal (2000) alone has a given
percentage error, is rather normal looking, biased at no more than the 1\% level,
and has $\sigma_{TF}$ $\approx$ 12\%. In fact, Sakai \etal have produced
four realizations of the TF H$_0$ measurement at different wavelengths, 
obtaining 74 \hunit
from an I band calibration and 67 \hunit from an H band calibration.
The chance that this 10\% discrepancy would occur by chance, especially
when the same calibrator distances have been assumed in both cases, is small.
Sakai \etal consider systematics in the linewidths, I band extinction
corrections, and H band aperture/diameter ratios as possible contributors
to this discrepancy. Uncertain homogeneity in galaxy diameters leads to
lower weight for the H band result.

The other panels in Figure 2 show the SBF error distribution,
the SN error distribution, and the FP error distribution.
The narrowest is the SN error distribution (9\% $rms$, compared with
12\% in the other two cases). Figure 3 shows the covariance between
the SN and TF calibration errors, which occurs because a number of
the SNIa calibrators are also TF calibrators. Cepheid bearing galaxies
represent only a selected sample (or in the FP case, merely neighbors)
of the population of galaxies to
which the calibration we have derived is applied. 
Stellar population effects are calibrated empirically in the case of SBF,
but we make no allowance for parameters (beyond decline rate and reddening),
which may still remain hidden in the SN case.

\section{Combining The Constraints}

The results of the previous section will aid us in optimally combining
the four secondary distance indicators. We can compare $<H_0>$ from
a straight mean of the four measurements and compare this with
a weighted mean. We choose weights for H$_0^{TF}$ 
and H$_0^{SBF}$ which are the inverse of $\sigma^2_{TF}$ and $\sigma^2_{SBF}$,
respectively. Effectively, this 1.5 times weights the SN distance indicator
relative to the other three.

Combining H$_0^{TF}$ = 71 $\pm$ 4  (random) $\pm$ 7 (systematic) (Sakai \etal 2000) 
with  H$_0^{SBF}$ = 69 $\pm$ 4 $\pm$ 6 (Ferrarese \etal 2000a), 
and  H$_0^{FP}$ = 78 $\pm$ 8  $\pm$ 10 (Kelson \etal 2000), 
and  H$_0^{SNIa}$ = 68 $\pm$ 2 $\pm$ 5 (Gibson \etal 2000), 
we obtain H$_0$ = 71 $\pm$ 6 \hunit, without the weighting
influencing the outcome a great deal.
The error distribution for the combined constraints is shown
in Figure 4. The width of this distribution is $\pm$9\% (1$\sigma$).

A dimension which each of the error distributions shares is dependence
on the assumed distance of  the LMC. This is illustrated in Figure 5,
which shows the 67\% probability contours for each of the four
secondary distance indicators. The only comparable distance indicator
in the local volume which does not depend on the LMC distance
(but is still influenced by SN1987A), is
the Expanding Photospheres Method applied to supernovae of Type II.
This yields H$_0$ = 73 $\pm$ 12 \hunit (Schmidt \etal 1994)
at 95\% confidence.

If we adopt Figure 1 as the probability distribution of the distance
of the LMC, the error distributions broaden and reflect the skew seen
in Figure 1. The combined constraint is shown in Figure 6. The uncertainty
in H$_0$ grows to 12\% and the bias, the amount by which H$_0$ is underestimated
through our assumption of a 50 kpc distance
becomes 4.5\%. It is likely that Figure 1 exaggerates the probability
of LMC distance moduli as low as 18.1 mag, as it weights recent
estimates based on the brightness of the ``red clump" almost as highly 
as it weights Cepheids. A critical literature review on the LMC distance
is provided by Freedman \etal (2000). Expressing our 
result as a self-contained experiment, we obtain H$_0$ = 
3.5 $\pm$ 0.2 \kms per LMC distance.

We conclude that the expansion rate within the area mapped by the
secondary distance indicators we have calibrated
is 71 $\pm$ 6 \hunit. The distribution of galaxies in 
Figure 7 renders this result relatively immune to
a low amplitude bulk flow of the sort detected by Giovanelli \etal (1998).
Similarly, velocity perturbations due to a Comacentric bubble or a Local
Void (Tully \& Fisher 1987) would tend to generate a dipole in Giovanelli's
results, which is not seen, at least in the Arecibo sky sample available
to date.

Finally, we note that adoption of the metallicity dependence of the 
Cepheid PL relation described in $\S$3 reduces the combined H$_0$
by 4\% to 68 $\pm$ 6 \hunit.

\section{Future Work}

A large scale, locally centered bubble would require that this local H$_0$ be 
corrected for the density anomaly. 
Tammann (1998) and Zehavi \etal (1998) have estimated that this amounts to a few percent,
and this deserves careful evaluation focussed on the volume sampled in the
accompanying papers.
Truly $large$ scale density perturbations are correspondingly unlikely
(Shi \& Turner 1998).

To complete the Key Project, we intend to examine this matter and several
other limitations of current work, which have been identified here and in the
accompanying papers. These include the limited LMC Cepheid PL relation,
the excessively large uncertainty in the photometric calibration we have
adopted for WFPC2, and the comparison of results from this classical approach 
to the Extragalactic Distance Scale with recent progress in the analysis
of gravitationally lensed quasar time delays and the x-ray gas in
rich clusters of galaxies (the Sunyaev Zeldovich effect). This work is
in progress (Freedman et al 2000). The Key Project's error analysis will
also be developed in more detail than we have presented here.

These additional steps should secure the Key Project's goal -- a 10\%
Hubble Constant -- to a higher level of confidence than the 1$\sigma$
level reported here.
\acknowledgments

The work presented in this paper is based on observations with the NASA/ESA
Hubble Space Telescope, obtained by the Space Telescope Science Institute,
which is operated by AURA, Inc. under NASA contract No. 5-26555.
Support for this work was provided by NASA through grant GO-2227-87A from
STScI. SMGH and PBS are grateful to NATO for travel support via a Collaborative
Research Grant (960178).  Collaborative research on HST data at Mount
Stromlo was supported
by a major grant from the International S \& T program of the Australian
Government's Department of Industry, Science and Resources.
LF acknowledges support by NASA through Hubble Fellowship grant
HF-01081.01-96A. SS acknowledges support from NASA through the 
Long Term Space Astrophysics Program, NAS-7-1260.
We are grateful to the Lorentz Center of Leiden University for its
hospitality in 1998, when this series of papers was planned.
We would like to thank Riccardo Giacconi for instigating the HST Key Projects.
\clearpage

\appendix{\bf Appendix A. The Local Flow Field}

The model outlined in $\S$2 employs a five step procedure to convert
heliocentric velocities to velocities characteristic of the expansion
of the Universe.

1) Correction of the observed heliocentric velocity of our objects to the centroid
of the Local Group.  We use here the Yahil, Tammann and Sandage (1977)
prescription (YST) for consistency, but note that use of other prescriptions 
(e.g. the IAU 300 sin(l)cos(b)) generally does not make a large difference
beyond halfway to Virgo.
The YST correction to the Local Group centroid is
$$V_{LG} \ = \ V_H - 79 \cos(l) \cos(b) + 296 \sin(l) \cos(b) -36 \sin(b) 
\qquad (A1).$$
As indicated in $\S$2, we set $V_0 \ = \ V_{LG}$.

2) Correction for Virgo infall.  Note that the Virgo cosmic velocity is
derived by correcting the observed heliocentric velocity (Huchra 1995)
to the LG centroid, for  our infall velocity and for its infall into the GA.
Note, again, that the correction for Virgo infall includes {\it two} components,
the change in velocity due to the infall of the object into Virgo plus
the vector contribution  due to {\it the Local Group's} peculiar
velocity into Virgo.  That term is just $V_{fid} \cos(\theta_v)$

3) Correction for GA infall as in 2).

4) Correction for Shapley supercluster infall. The correction adopted is set
so that it reproduced the amplitude of the CMB dipole as V $\longrightarrow~
\infty$.

5) Correction for other concentrations as necessary.

\noindent Since we have set the solution to be additive, the
final corrected ``Cosmic''  velocity w.r.t. the LG is then

$$V_{Cosmic} \ = \ V_H + V_{c,LG} - V_{in,Virgo} - V_{in,GA} - V_{in,Shap} -... \qquad (A2),$$

\noindent where $V_H$ is the observed heliocentric velocity and
$V_{c,LG}$ is the correction to the Local Group centroid described above.
Note that the STY correction to the Local Group centroid is {\it not}
the same as the IAU correction, so some of the models and assumptions
made in earlier Virgo flow fits have to be modified.  We have
used the YST assumption primarily because it is what was used to derive the
corrected CMB dipole.

For our initial attempt at a detailed flow field correction, we include
just three attractors,  the Local Supercluster, the Great Attractor and
the Shapley Supercluster.  The 
parameters we use for the attractors are given in Table A1 and are taken
(and estimated) from a variety of sources including AHMST, Han (1992), Faber 
\& Burstein (1989), Shaya, Tully \& Pierce (1992)  and Huchra (1995).   
For simplicity, we also assume $\gamma$ = 2.  For this first cut
model, we assume an infall into Virgo of 200 \kms at the LG, an infall
into the GA of 400 \kms and an infall into Shapley of 85 \kms.  These
numbers give good agreement with the amplitude of the CMB dipole, but
with only these three attractors, the direction of maximum LG motion
is 27 degrees away from the CMB direction.

\pagebreak
 
\appendix{\bf Appendix B: The Photometric Zeropoint}

The status and calibration of the Wide Field Planetary Camera 2
(WFPC2) has been reviewed by Gonzaga \etal (1999), who find that
photometric accuracies of a few percent are routinely possible.
The baseline photometric calibration for WFPC2 is given by Holtzman \etal
(1995). The standard calibration for papers IV to XXI in the Key Project
series is that of Hill \etal (1998), and accounts for the principal
systematic CTE effect, the so-called long vs. short
exposure effect (Wiggs \etal 1999). Photometric stability has been
satisfactory over the duration of the project with fluctuations of 
$\simlt$2\% or less peak-to-peak over 4 years at the wavelengths
observed here (Heyer \etal 1999).

Images obtained with the CCDs in WFPC2 are known to be subject to
charge loss during readout, presumably due to electron traps in
the silicon of the detectors.  Approximate corrections for this
charge loss have been published by Whitmore \& Heyer (1997) and
Stetson (1998), but these corrections are based on comparatively
short exposures of comparatively bright stars, so the observations 
exhibit a comparatively narrow range of apparent sky brightness.
Furthermore, the zero points of the WFPC2 photometric system are primarily
determined from comparatively bright stars, since those are the ones for
which the ground-based photometry is most reliable.  Since the amount of
charge lost from a stellar image appears to be a function of both the
brightness of the star image and the apparent brightness of the sky, these
dependences must be quite well determined to enable reliable extrapolation
from bright stars as observed against a faint sky (standard stars in
relatively uncrowded fields in short exposures) to faint stars observed
against a bright sky (distant Cepheids project against galactic disks in
long exposures).  The study of Stetson (1998) was intended to provide
that extrapolation, based upon comparatively short exposures of the
nearby globular cluster $\omega$~Centauri, and both short and long
exposures of the remote globular cluster NGC~2419.  These data
seemed to yield consistent charge-loss corrections and zero points,
but various tests suggested that there remained some uncontrolled
systematic effects which might amount to of order $\pm$0.02 mag
or so in each of the $V$ and $I$ filters.

Improved correction for Charge
Transfer Efficiency effects in the WFPC2 CCDs has been presented by
Stetson (1998). Presumably because of different illumination levels,
correction of our photometry affects the V magnitudes and the I magnitudes
differently. In the mean, and based on the reference stars photometry
published in papers IV to XXI, distance moduli on the Stetson (1998) system
are 0.07 $\pm$ 0.02 mag closer than on the Hill \etal system.

In attempting to improve upon this situation, Stetson (work in progress)
has added ground-based and WFPC2 data for the nearby globular cluster M92
(= NGC 6341) to the solution.  The WFPC2 data for M92 are
intermediate in exposure time between those for $\omega$ Cen and the short
exposures of NGC 2419 on the one hand, and the long exposures of NGC 2419
on the other.  As in Stetson (1998), the $\omega$ Cen, NGC 2419, and M92
data were all combined into a single solution to determine the optimum
coefficients relating the amount of charge lost from a stellar image to its
position on the detector, its brightness, and the surface brightness of the
local sky.  When these corrections are applied to determine the optimum
photometric zero points from the data for each globular cluster, it is
found that the results for $\omega$ Cen and NGC 2419 are consistent, as
before, but the zero points implied by the M92 data are substantially
different:  $+0.054\pm0.003\,$mag in $V$, and $-0.038\pm0.003$ in $I$.
Adoption of a zeropoint based on M92 would move the Key Project galaxies 
0.14 mag closer than the Hill \etal reference point.

On the other hand, Saha (in preparation) finds different results from analysis
of Cycle 7 calibration data. He has determined that CTE correction will 
yield Cepheid colors bluer by $\approx$0.02 mag, corresponding to 
distance moduli $more~distant$ by 0.05 mag. 

Given these uncertainties, we continue to adopt the 
Hill \etal (1998) calibration, but we note that CTE effects render our
distance moduli more uncertain than we have previously estimated.
The modulus uncertainty adopted here is $\pm$0.09 mag.
That makes Stetson's M92 results a 1.5$\sigma$ anomaly.
Physically, the M92 results seem anomalous, since CTE correction should
be  intrinsically grey. Further investigation of the zeropoint for the Cepheid
photometry database is required.

\pagebreak

\pagebreak

{\large \bf Figure Captions}

Figure 1. Distribution of published LMC distance moduli from the literature. Values from 1983--1995 are from the review by Westerlund (1996). Values
up to the end of 1998 have been collated by Freedman (1999).

Figure 2. The distribution of uncertainties in H$_0$
for each of the four secondary distance indicators calibrated and applied 
in the Virtual Key Project.

Figure 3. Percentage error contours for the supernova and Tully-Fisher
measurements of H$_0$. The outer contour encloses 80\% of the realizations.

Figure 4. The uncertainty distribution for the combined constraints on H$_0$.

Figure 5. The Key Project calibration constrains H$_0$ in four ways,
but these are each, in turn, dependent on the assumed distance of the
Large Magellanic Cloud which provides the reference Cepheid PL relation
for the project.

Figure 6. The corresponding distribution calculated from the probability
distribution of LMC distances in Figure 1.

Figure 7. The distribution of secondary distance indicators in projection
on the supergalactic plane. Open circles: TF clusters from Sakai \etal (2000);
solid circles: SBF clusters from Ferrarese \etal (2000a); asterisks: SNeIa
from Hamuy \etal (1996); crosses: FP clusters from Kelson \etal (2000).

%\end{document}       %%%%%%%%%%%%%%%%% end here if you want no figures %%%%%%%

\clearpage

\begin{figure}[h]
\centering \leavevmode
%\vskip -5 truein
%\hskip -.7 truein
\epsfysize 7.5 truein
\epsfbox{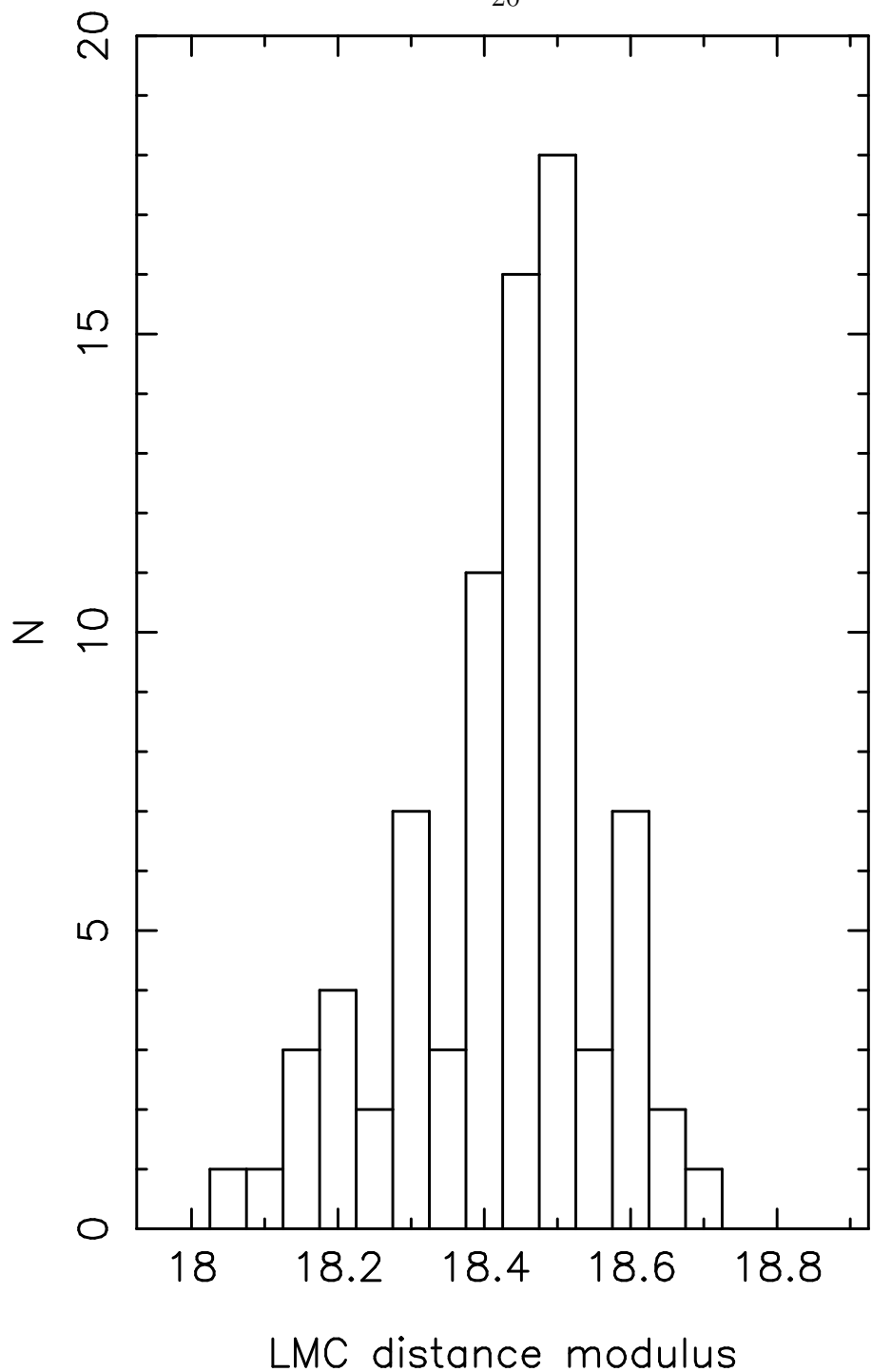}
%\vskip -.4 truein
\caption{ {\it Distribution of published LMC distance moduli from the literature. Values from 1983--1995 are from the review by Westerlund (1996). Values
up to the end of 1998 have been collated by Freedman (1999).
 }}
\end{figure}

\clearpage
\begin{figure}[h]
\centering \leavevmode
\vskip .5 truein
\hskip -.7 truein
\epsfysize 9 truein
\epsfbox{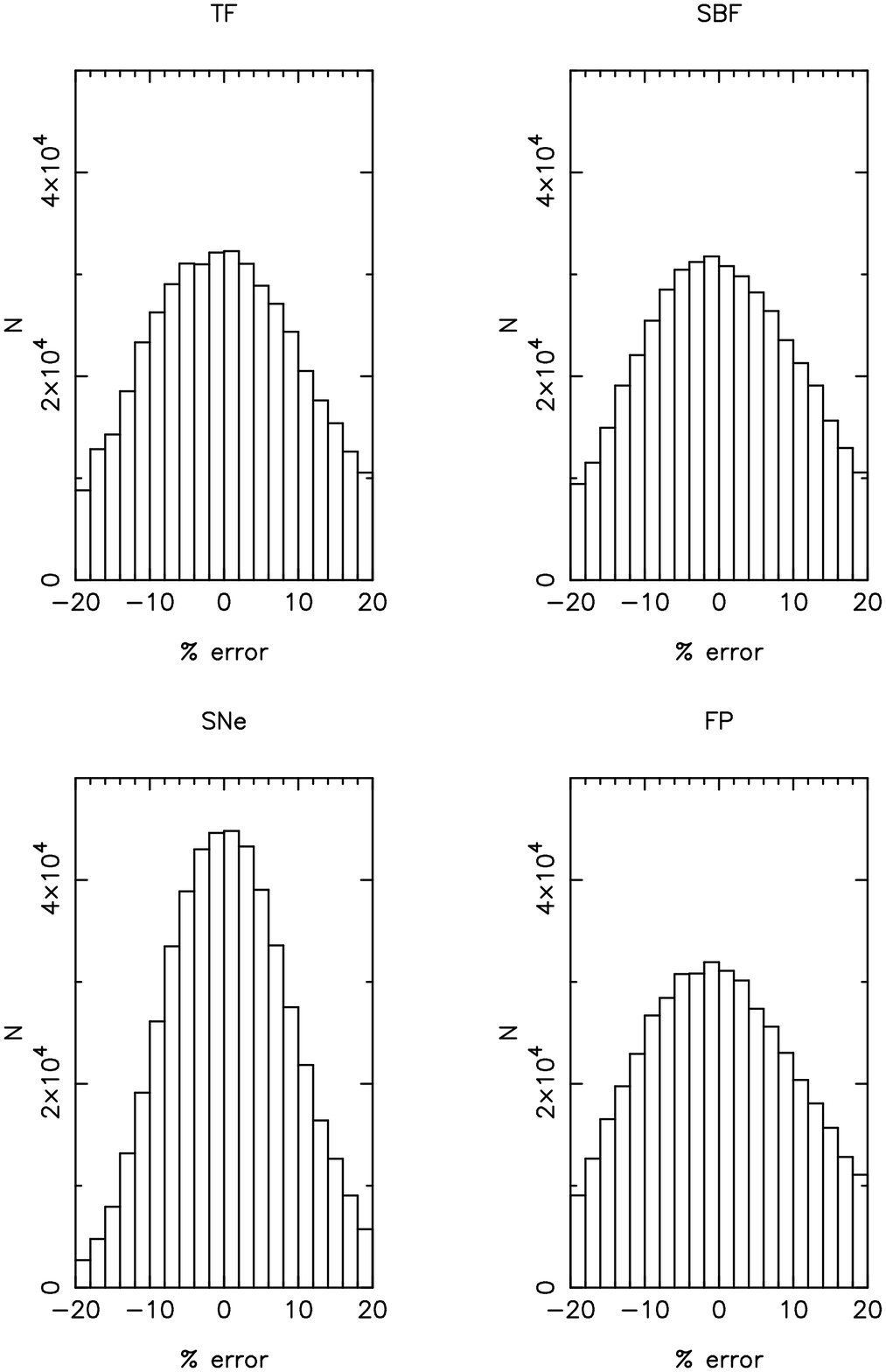}
\vskip -.4 truein
\caption{ {\it abcd The distribution of uncertainties in H$_0$ in each
for each of the four secondary distance indicators calibrated and applied 
in the Virtual Key Project.
 }}
\end{figure}

\clearpage
\begin{figure}[h]
\centering \leavevmode
%\vskip -5 truein
\hskip -.7 truein
\epsfysize 4 truein
\epsfbox{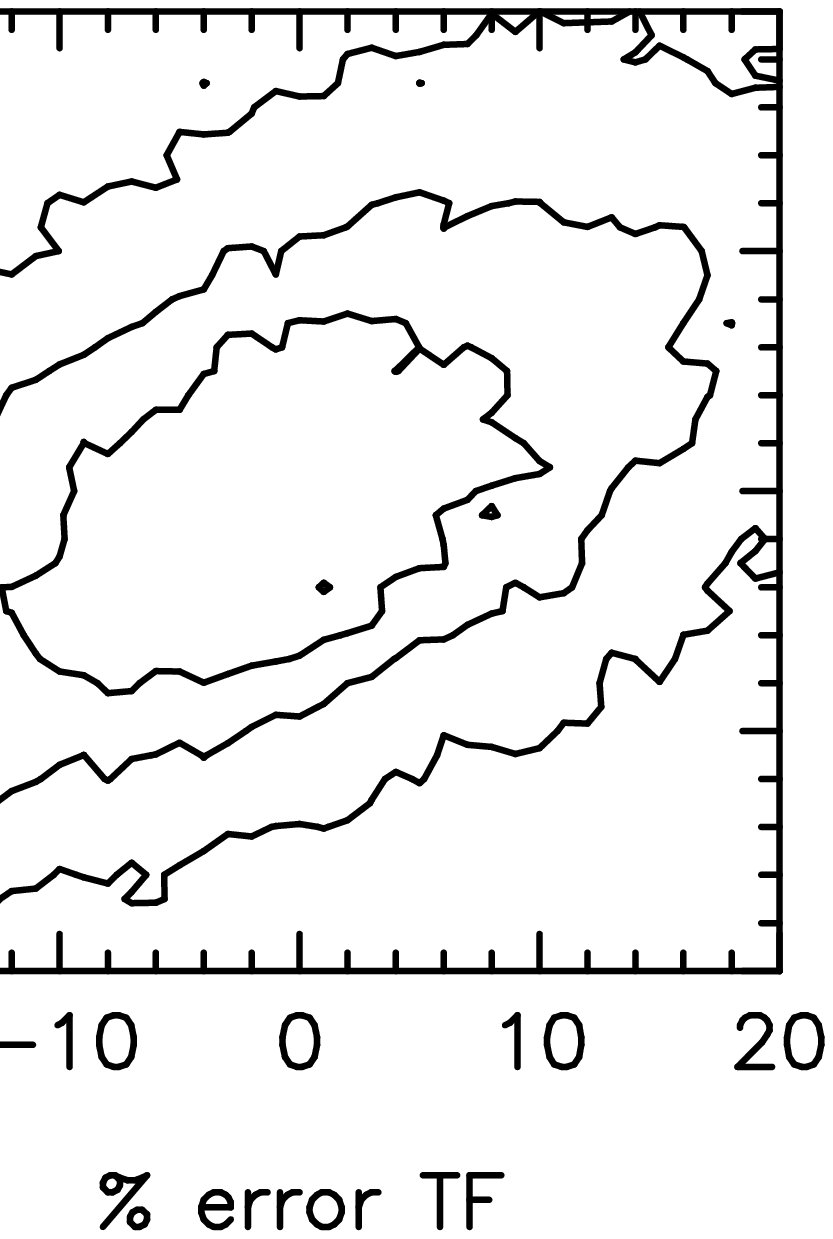}
%\vskip -.4 truein
\caption{ {\it Percentage error contours for the supernova and Tully-Fisher
measurements of H$_0$. The outer contour encloses 80\% of the realizations.}}
\end{figure}

\clearpage
\begin{figure}[h]
\centering \leavevmode
\vskip 2 truein
\hskip -.7 truein
\epsfysize 9 truein
\epsfbox{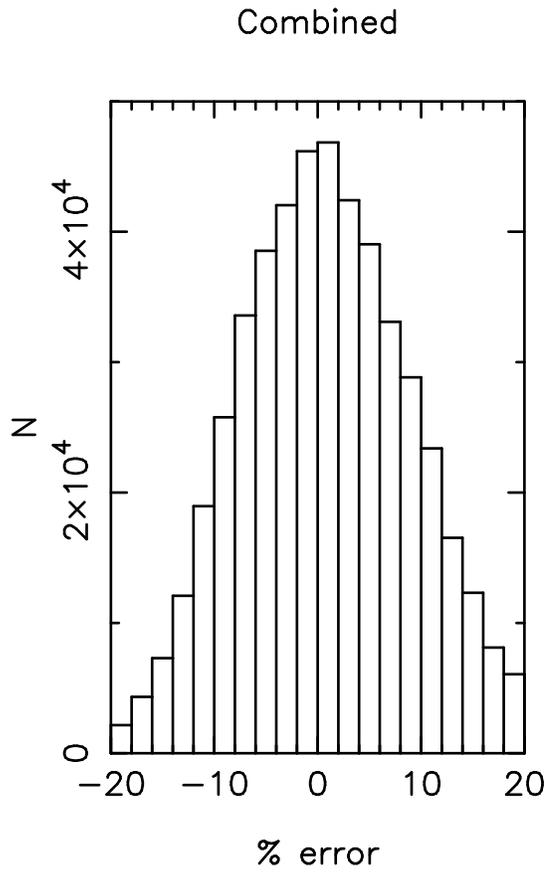}
\vskip -4 truein
\caption{ {\it The uncertainty distribution for the combined constraints on H$_0$.}}
\end{figure}

\clearpage
\begin{figure}[h]
\centering \leavevmode
\vskip 2 truein
\hskip -.7 truein
\epsfysize 9 truein
\epsfbox{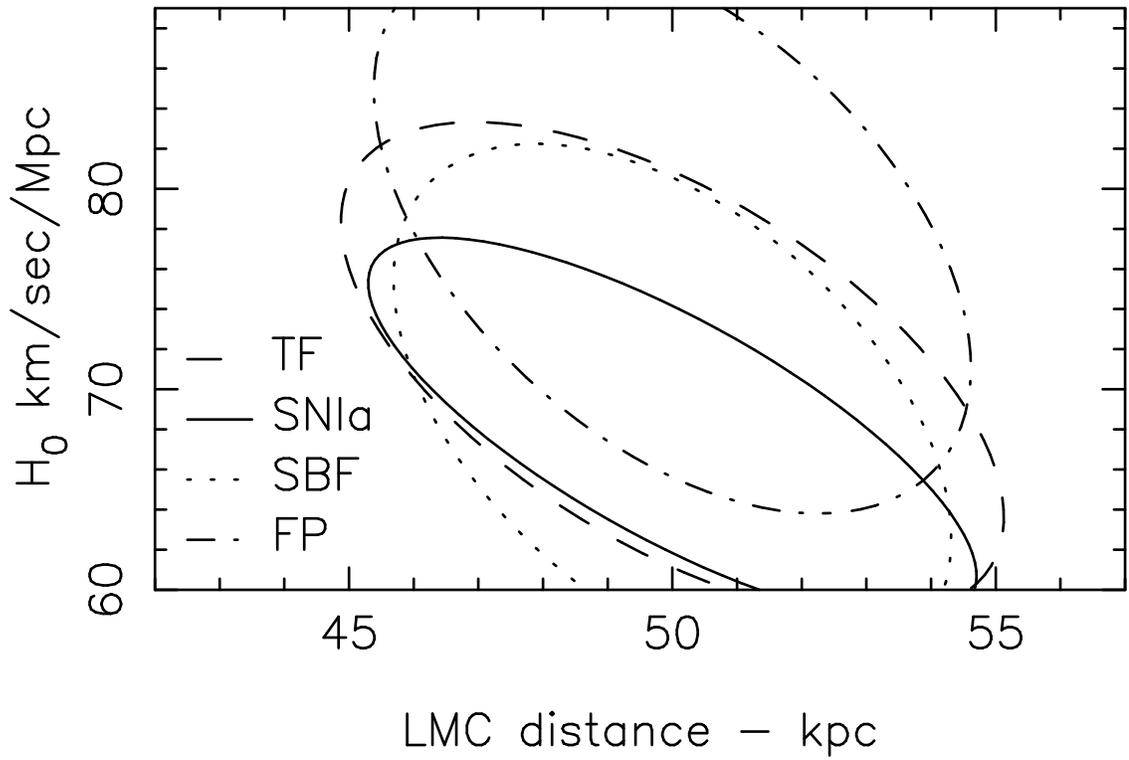}
\vskip -4 truein
\caption{ {\it The Key Project calibration constrains H$_0$ in four ways,
but these are each, in turn, dependent on the assumed distance of the
Large Magellanic Cloud which provides the reference Cepheid PL relation
for the project.}}
\end{figure}

%\clearpage
\begin{figure}[h]
\centering \leavevmode
\vskip 2 truein
\hskip -.7 truein
\epsfysize 9 truein
\epsfbox{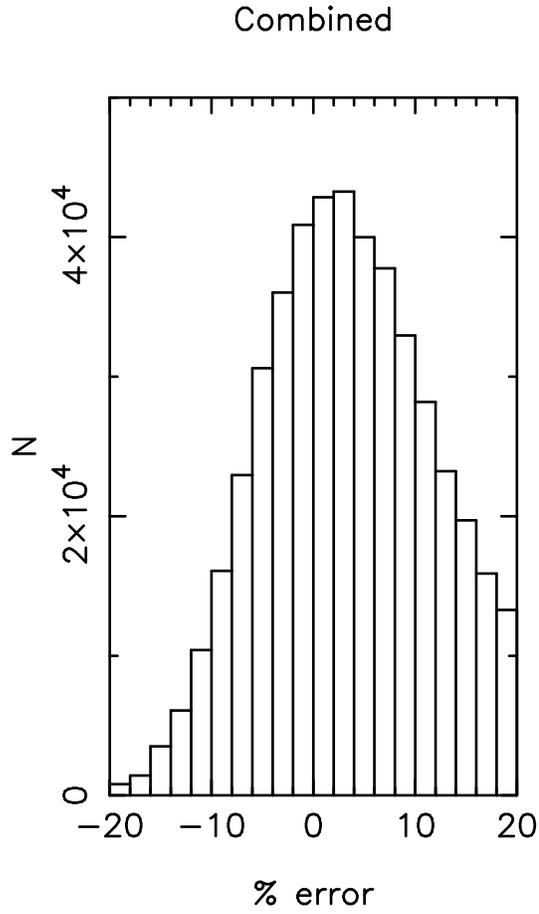}
\vskip -4 truein
\caption{ {\it The corresponding distribution calculated from the probability
distribution of LMC distances in Figure 1.}}
\end{figure}

\begin{figure}[h]
\centering \leavevmode
%\vskip -5 truein
%\hskip -.7 truein
\epsfysize 8 truein
\epsfbox{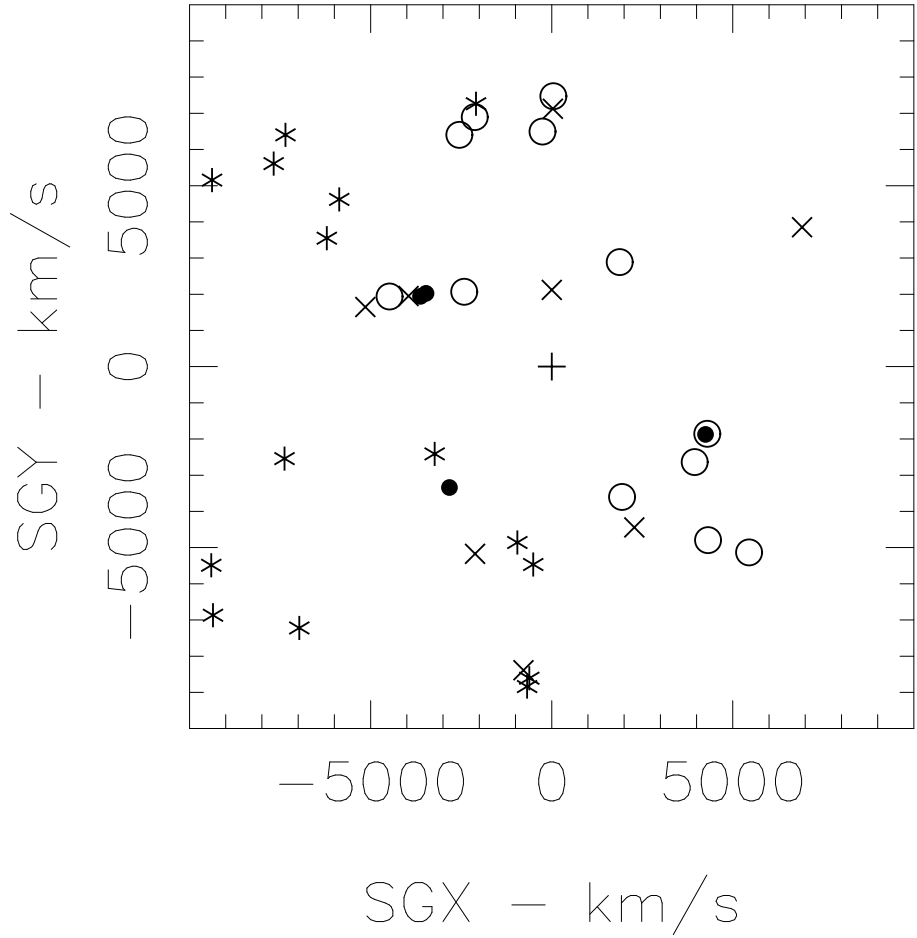}
%\vskip -.4 truein
\caption{ {\it The distribution of secondary distance indicators in projection
on the supergalactic plane. Open circles: TF clusters from Sakai \etal (1999);
solid circles: SBF clusters from Ferrarese \etal (1999); asterisks: SNeIa
from Hamuy \etal (1996); crosses: FP clusters from Kelson \etal (1999).
 }}
\end{figure}

\end{document}